\documentclass[sigconf]{acmart}
\copyrightyear{2020}
\acmYear{2020}
\setcopyright{acmcopyright}\acmConference[ICMR '20]{Proceedings of the 2020
International Conference on Multimedia Retrieval}{June 8--11, 2020}{Dublin, Ireland}
\acmBooktitle{Proceedings of the 2020 International Conference on Multimedia
Retrieval (ICMR '20), June 8--11, 2020, Dublin, Ireland}
\acmPrice{15.00}
\acmDOI{10.1145/3372278.3390705}
\acmISBN{978-1-4503-7087-5/20/06}
\usepackage{balance}

\AtBeginDocument{%
  \providecommand\BibTeX{{%
    \normalfont B\kern-0.5em{\scshape i\kern-0.25em b}\kern-0.8em\TeX}}}

\usepackage{float}

\begin{document}

\fancyhead{}
\title{Continuous Health Interface Event Retrieval}


\author{Vaibhav Pandey}

\affiliation{%
  \institution{University of California Irvine}
  \streetaddress{3209, Donald Bren Hall}
  \city{Irvine}
  \state{California}
  \postcode{92617}
}
\email{vaibhap1@uci.edu}

\author{Nitish Nag}
\affiliation{%
  \institution{University of California Irvine}
  \streetaddress{3209, Donald Bren Hall}
  \city{Irvine}
  \state{California}
  \postcode{92617}
  }
\email{nagn@uci.edu}

\author{Ramesh Jain}
\affiliation{%
  \institution{University of California Irvine}
  \streetaddress{3209, Donald Bren Hall}
  \city{Irvine}
  \state{California}
  \postcode{92617}
  }
\email{rcjain@uci.edu}

\renewcommand{\shortauthors}{Pandey, et al.}

\begin{abstract}
  Knowing the state of our health at every moment in time is critical for advances in health science. Using data obtained outside an episodic clinical setting is the first step towards building a continuous health estimation system. 
  \newline
  In this paper, we explore a system that allows users to combine events and data streams from different sources and retrieve complex biological events, such as cardiovascular volume overload, using measured lifestyle events. These complex events, which have been explored in biomedical literature and which we call \textbf{interface events}, have a direct causal impact on the relevant biological systems; they are the interface through which the lifestyle events influence our health.
  We retrieve the interface events from existing events and data streams by encoding domain knowledge using the event operator language. 
  The interface events can then be utilized to provide a continuous estimate of the biological variables relevant to the user's health state. 
  The event-based framework also makes it easier to estimate which event is causally responsible for a particular change in the individual's health state.
\end{abstract}



\maketitle

\section{Introduction}
Usually, people become aware of their changing health state when they perceive symptomatic signals. For example, fever or inflammation from an infection results in the signals of feeling unwell, prompting an individual to seek medical attention. While this has worked for infectious diseases, where the cause of the ailment was an external agent, it does not work as well for chronic diseases where the cause may be events in daily life that have persisted for years. These events slowly change people's health to an irreversible state, without symptomatic warning. The most widely spread disease burden in the world is cardiovascular disease, and factors that change our cardiovascular health state act without any perceivable signals \cite{Mozaffarian2008BeyondDisease.}\cite{Anderson1991CardiovascularProfiles}. While there are tests available to measure a person's risk for chronic diseases, they are usually performed episodically and only in a clinical setting. In the example of cardiovascular disease, the invasive test of choice would be an angiogram \cite{Tavakol2012RisksReview.}. Since these tests are so invasive, they cannot be performed frequently enough to have a constant understanding of the health of an individual. This highlights the need for using unobtrusively collected lifestyle data for estimating individual health parameters \cite{Nag2018}.

With the rising popularity of wearable, mobile, and IoT devices in all aspects of life, individuals are generating more lifelog data than ever before in human history. These sensors are capable of tracking our life the whole day, every day. Lifelogging systems do precisely this \cite{Oh2017}, and track events related to a person's actions (such as exercise or food consumed), or their physical environment (such as exposure to smoke from a fire). Capturing all such events and data streams allows us to enhance the quality of continuous health monitoring systems. Our location history throughout the day can be combined with publicly available pollution and environmental data for computing our exposome, which has been found useful in studying underlying causes of different diseases and assessing disease risk factors\cite{Misra2019TheEra.}\cite{Sille2020TheAssessment}\cite{Bessonneau2019MappingCancer}. There are IoT devices that can capture air quality and particulate matter concentration even when we are indoors, thus making our exposome even more accurate\cite{LaserBetter}. These user-generated data and event streams can be used to identify recurring user habits and lifestyle choices which impact their health in the long and short term.
The importance of multimodal health tracking and estimation systems is even more apparent in the current context of the widespread pandemic of COVID-19. There are ongoing research studies to build early-diagnosis systems for COVID-19 using multimodal data collected from a variety of wearable devices \cite{Detect}\cite{UCSFStudy}. Such systems can have an immense impact in our fight against this pandemic by diagnosing individuals before they become symptomatic and significantly reducing the community spread of the disease.

While these devices do capture a large amount of data about events happening in our lives, they typically stay in their silos. It is necessary to combine all such events and data streams to enable multimodal applications to perform effective health estimation and guidance.
We may want to combine events or data streams from different sources to identify more complex events that have a deterministic relationship with health parameters. 
Such events act as an interface between lifestyle events and biological systems; hence we are calling them \textbf{interface events}. These events determine the mechanism through which lifestyle events impact our biology, and variations in these interface events could lead to entirely different outcomes for similar lifestyle events.

\begin{figure}[t]
  \centering
  \includegraphics[width=\linewidth]{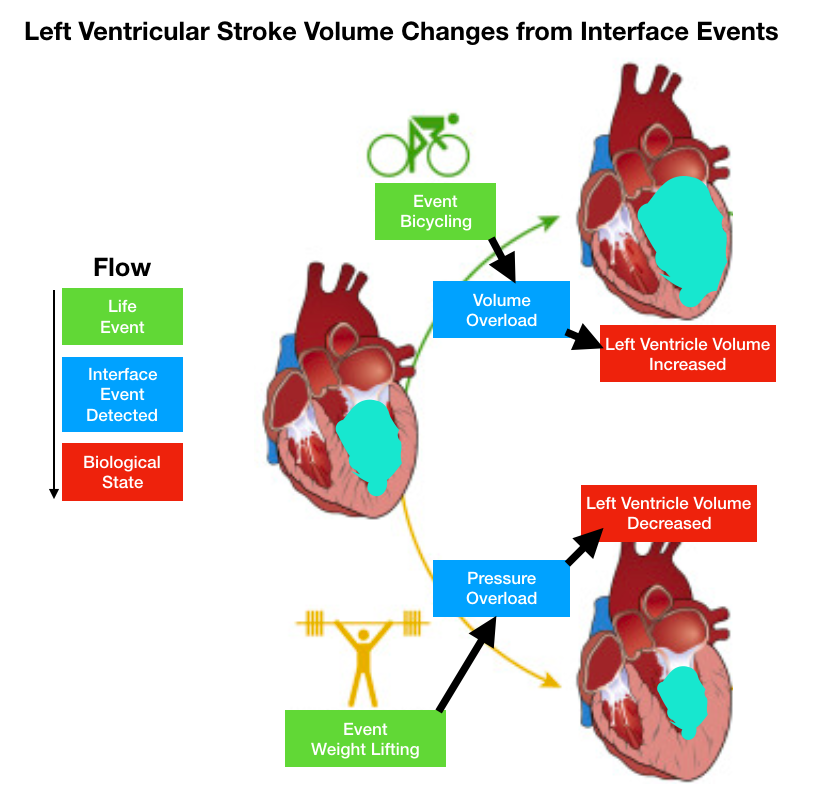}
  \caption{This figure shows that detecting relevant interface events is necessary to estimate the physiological state of an individual from their lifelog information. Here two examples of life events are given that cause opposite outcomes in cardiac adaptation. Both events at a crude level can be considered exercise, but precisely extracting the interface event is critical if we are to use the lifelog information to estimate evolving health states.}
  \Description{Interface Events Interacting with Heart Biology}
  \label{fig:heart}
\end{figure}

This can be observed when considering the long-term cardiac effects of different types of physical activities. Physical activities can strain our heart muscles in two ways, cardiovascular volume overload and pressure overload.
Cardiovascular volume overload is a biological event where the heart is required to pump large volumes of blood through the circulatory system. This event would cause structural changes in the heart in the form of eccentric hypertrophy, where the left ventricular chamber volume increases. The volume overload interface event could be caused by a variety of exercise events such as cycling, jogging, soccer, dance, or hiking. Other exercise events such as weight lifting, sprinting, bouldering or even bowel constipation would cause pressure overload in the heart. These events would cause concentric hypertrophy of the heart\cite{Muller2013DifferencesHypertrophy}\cite{Mihl2008CardiacAthletes}\cite{deSimone2004ConcentricDifference}. This is illustrated in Figure \ref{fig:heart}. In this case, we see that similar lifestyle events can have a vastly different impact on our heart, and it is important that we find the cardiovascular volume and pressure overload events if we want to find the impact of any exercise on our heart.

Another simple example of an interface event would be an acute stress response from the body. This could be caused by a public speaking situation, fear of a threat, an overload of work at the office, or an emotional situation. All these events are quite varied, yet a primary influence on the physical biology will be through the stress response of releasing cortisol into the bloodstream and an increase in blood pressure \cite{Kirschbaum1993TheSetting}.
\newline
In our work, we demonstrate a system that allows us to aggregate data from different sources to enrich an existing event or create new interface events. We can find lifestyle events that cause changes in health by finding the interface events associated with lifestyle events.
We are trying to answer the following  research questions (RQ) in this work:
\begin{itemize}
    \item RQ1. How can we use different sources to identify a specific interface event?
   \item RQ2. How can we retrieve an interface event continuously over time using its definition? 
    \item RQ3. How can we convert a real-world query to relevant interface events and identify the required attributes and patterns?
\end{itemize}

\section{Related Works}
\subsection{Daily Events Aggregation: Lifelogs}
Aggregating and recognizing events in our daily lives is a popular problem in the multimedia community. This problem has been termed "Lifelogging", which is explained as "a phenomenon whereby people can digitally record their own daily lives in varying amounts of detail, for a variety of purposes"\cite{Gurrin2014LifeLogging:Data}. Lifelogging is the first step towards recognizing various daily activities and how they define user's behavior.
Lifelogging applications collect a variety of data streams about an individual, which has the potential to offer unique insights into human behavior. There are many visual lifelogging applications and projects which aim to understand the user's life and activities by using the images taken by a wearable camera (E.g., GoPro) over a long period and the data collected could be used to identify events and daily activities happening in user's life for varied purposes\cite{Wang2018ComputerSemantics}\cite{ElAsnaoui2017ADatasets}\cite{Dobbins2017DetectingLiving}\cite{Wang2013UsingActivities}.

The daily activity recognition could be further enhanced by merging the visual log with other multi-sensory data that can be collected using smartphones, wearable devices, and different IoT systems \cite{Alameda-Pineda2019MultimodalIntroduction}. Some applications attempt to accomplish this task without using the visual logs to make the logging process more unobtrusive\cite{Oh2017}.

\subsection{Event retrieval frameworks}
People tend to view their time and experiences as different events that occurred in their lives. Even when we do remember an event from our lives, rarely do we remember the experiential information associated with it. Thus events provide a very natural abstraction over the underlying data streams generated by various systems and are easily understood by people. They also provide a very convenient mechanism to understand the behavior of a system even though the nature of the underlying data streams may change.
With the rising popularity of smartphones, cameras, wearable devices, and IoT sensors, people are generating a humongous volume of data. Thus event recognition and retrieval are critical to utilize and manage the growing volume of data effectively. 
\subsubsection{Data-driven retrieval} A large amount of data on the web is unstructured (E.g., Images, Videos) when it comes to extracting semantic value from it. Data-driven retrieval is appropriate in such situations, where a large volume of data is available, but the exact relationship between available data and the events is not well-defined or structured. It is especially the case for event retrieval from image, video, or audio streams. We would typically see data-driven multimodal event recognition and retrieval in surveillance systems where we want to recognize a set of predefined events from multimodal data streams (audio and video), and there is a large amount of tagged data available for training machine learning models. \cite{Atrey2006} defines a multi-level feature extraction and training methodology to detect events from audio streams. \cite{Valenzise2007ScreamSystems} describes a system to recognize scream and gunshot events from audio streams. \cite{KumarAtrey2006InformationSystems} describes a system to utilize multiple modalities to detect and retrieve an event. A lot of these multimedia event detection systems adopt a two-level process, first is feature generation at a clip level that is in-turn generated by aggregating frame-level features. The second step is using these features to train a classifier. This makes the system "static" as it uses only the local temporal features. With recent developments in deep learning, Recurrent Neural Networks have gained popularity for recognizing events in any stream of data and also allow us to consider an extended range of temporal information \cite{Wang2016Audio-basedNetworks}. \cite{Jiang2017AutomaticRNN} and \cite{Ramanathan2016DetectingVideos} attempt to recognize events in multi-person videos in a sports setting (soccer and basketball, respectively) using RNNs. 
\subsubsection{Knowledge-driven retrieval} In many problems, there exists an underlying syntactic or semantic structure in data streams. This is especially true for natural language processing, where the data follows the grammatical rules of the language. Knowledge-driven retrieval is appropriate in such situations, where we know the composition of events in terms of other events or patterns observed in data streams. This kind of structure allows us to specify the events as patterns of other events or patterns in data streams. This has been demonstrated in \cite{Hung2010WebLabeling}, \cite{Nishihara2009EventBlogs}. \cite{Hung2010WebLabeling} utilizes lexico-syntactic patterns related to an event to identify and extract the sentences about the event from web search results. These sentences are then used to build a commonsense knowledge base about the event. \cite{Ijntema2012AText} describes a lexico-semantic pattern language to extract event relationships from a text, which is then later used to learn ontology instances from the text that help domain experts with ontology population process. In all of these instances, a pattern description language is used to encode the domain-knowledge sourced rules to parse the incoming data and retrieve the events from text data. We have adopted a similar approach in multimodal streams, where an event pattern language is used to define complex events using directly measured events or events recognized from a subset of data streams, and the patterns described by these rules are obtained from medical literature.

\section{Events Aggregation Framework}
There are various publicly available applications and devices that capture and measure different user activities. We can track our sleep events and measure sleep quality using wearable devices (e.g., FitBit, Garmin, Apple Watch) or smartphone applications (e.g., Sleep Cycle). We have applications like MyFitnessPal that allow us to create a log of our meals, along with the nutritional content of the food items. There are applications to track workouts and athletic activities that collect data at a higher resolution than daily life activities (e.g., Strava, Training Peaks) and measure our performance metrics for those activities in addition to physiological parameters such as heart rate.
Many of these applications collect data without user intervention, which is one of the reasons for the widespread adoption of such systems and allows us to build future applications on top.

Aggregation of events and data of different modalities and from different sources is necessary to fully utilize different ubiquitous systems tracking and measuring different aspects of our lives\cite{Pandey2018}. We have to combine the incoming data in one log for the individual, which can then be utilized to identify their habits and how those impact the user's life. 
Building the Personal Event Chronicle is an active research topic, and many researchers are working on systems to recognize and combine daily life events from different sources and sensors in an unobtrusive manner. \cite{Oh2017} describes a system that utilizes a smartphone to collect various data streams and segment an individual's day into widely recognized life events \cite{Kahneman2004AMethod}. 
\newline
In this work, we propose to use any such system which recognizes a subset of life events and aggregate the output of those systems to recognize events that directly impact user's health in one or more ways.
We view every system as a source for a stream of events, and each event may also have a set of data streams associated with it, which provide additional information about the event. For example, a running event captured using a wearable device may have been recognized using the corresponding accelerometer data and has associated heart rate and GPS position streams, which provide information about the event such as calories burned, altitude change, and average pace during the event.

\begin{figure}[h]
  \centering
  \includegraphics[width=\linewidth]{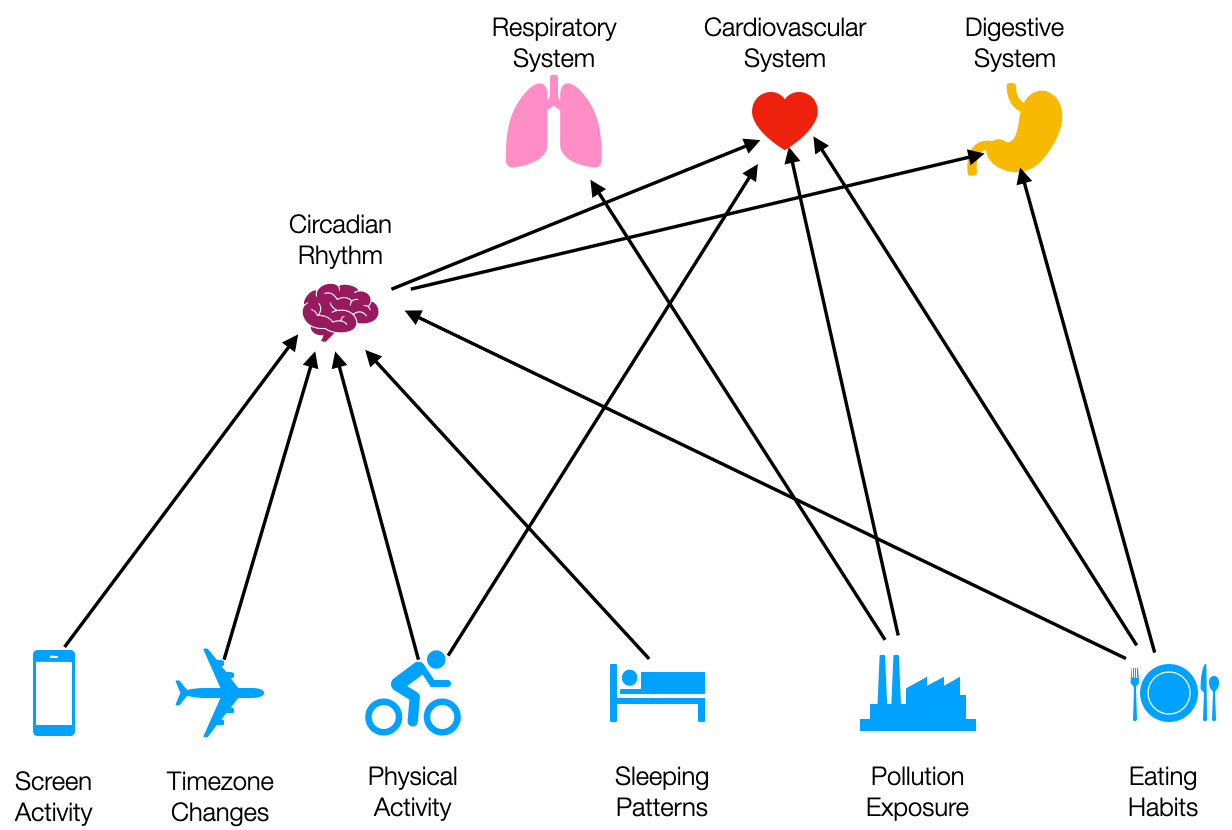}
  \caption{PM2.5 exposure has been linked with heart failure and death \cite{Shah2013GlobalMeta-analysis}\cite{Sun2010CardiovascularExposure}\cite{Kampa2008HumanPollution}. 
 Regularly sleeping and eating at abnormal circadian times is associated with increased risk for obesity, diabetes, and cardiovascular disease\cite{Scheer2009AdverseMisalignment}. Lifestyle events such as time zone changes, light exposure, and exercise and sleeping habit variations can cause circadian disruptions. 
 }
  \Description{Lifestyle event and health parameter relationship.}
  \label{fig:event-relationship}
\end{figure}

\section{Knowledge Driven Event Retrieval}
\begin{figure}[h]
  \centering
  \includegraphics[width=\linewidth]{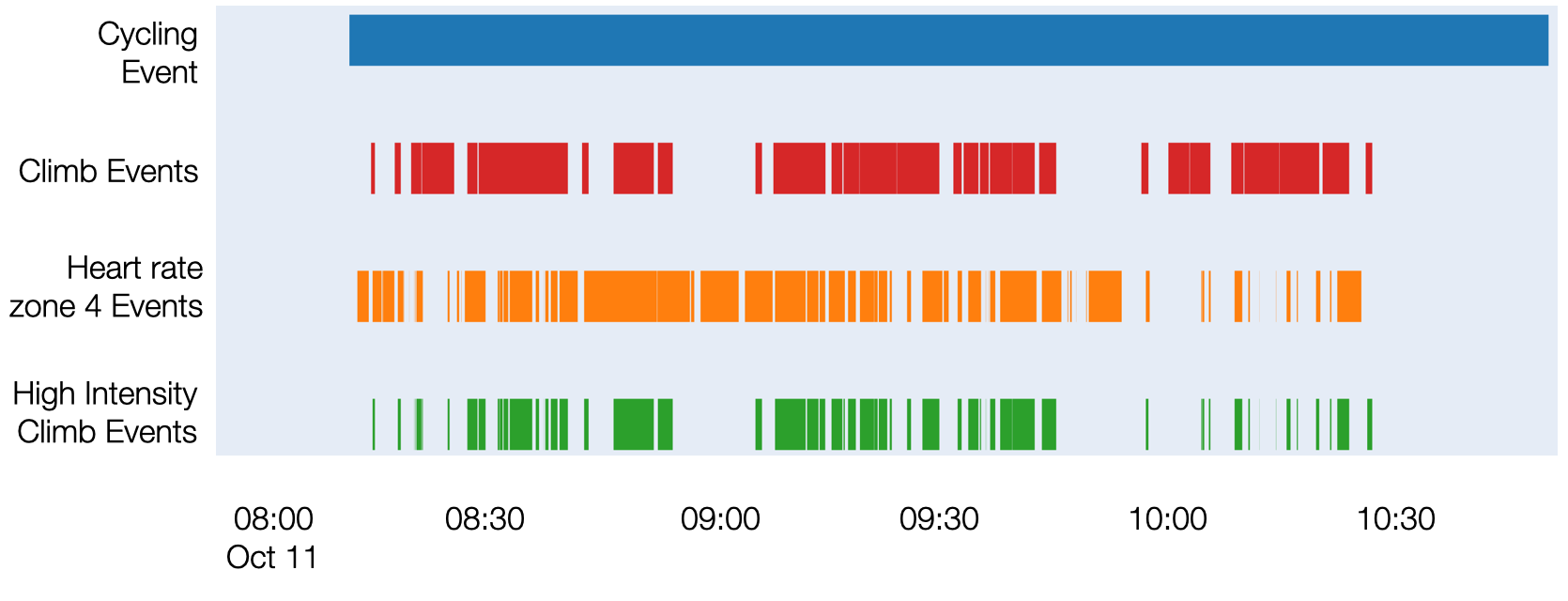}
  \caption{This figure shows the application of AND operator to combine events. We are trying to detect interface events where the heart rate of the individual is in zone 4 (170-190 bpm) while they are climbing a slope on a bicycle.}
  \Description{Climb Event Plot.}
  \label{fig:climb-event}
\end{figure}

Various events in our daily lives have a direct long-term and short-term impact on our health. Every time we eat, sleep, or walk up a set of stairs, our health state changes imperceptibly. Over time, these changes accumulate and may present themselves as chronic diseases. In order to provide health navigation for individuals\cite{Nag2019ALife}, we need to recognize all such events that are relevant for determining a user's health state. We have adopted a knowledge-driven approach to recognize and retrieve those events. We are calling these \textbf{interface events} as these act as an interface between daily events and biological systems.
\newline
A lot of these interface events have been recognized in the medical literature. For example, PM2.5 exposure and circadian rhythm disruption are caused by different lifestyle events such as outdoor exercises, changes in sleeping and eating habits, and screen activity and impact the functioning of various biological systems (Figure \ref{fig:event-relationship}).
Similarly, links between diet and various chronic diseases such as type II diabetes, hypertension, and ASCVD has been well established in the literature. 
Thus, we must utilize the existing knowledge to guide the retrieval of these interface events from user-generated data.
\newline
The interface events can be specified as transformations and combinations of existing events and data streams. These transformations can be as varied as applying a threshold on a data stream (e.g., binning heart rate in different ranges) to recognizing events and actions from a video stream (e.g., fall detection).
\newline
We currently use a subset of the event pattern language proposed in \cite{Jalali2016} to describe the interface events. We have implemented a subset of the event operators described in \cite{Jalali2016} specifically NOT $(\neg)$, AND$(\wedge)$, OR$(\vee)$ and DELAY$(\delta)$ operators, which can be used in conjunction with user-defined event detection operators to describe complex events. For example, if we want to identify events where the heart rate of the individual is above 120 bpm, and PM2.5 concentration is above 10 $\mu$g/$m^3$ of air, then we would need first to create a PM2.5 stream using their location stream. We could then define these exposure events using the formulation given below. Figure \ref{fig:pm25-event} shows some of the retrieved events.
\newline
\begin{displaymath}
  ExposureEvent := (Heartrate > 120) \wedge (PM2.5 > 10)
\end{displaymath}
\newline
These event operators could be combined with user-defined event recognition operators allowing us to utilize more data streams and recognize more complicated events. For example, we define a climb-detection operator for the altitude data stream. This operator detects time intervals when the user is climbing up a slope. The detected events could be combined with cycling events and events where the person's heart rate is above 170 to find out high-intensity uphill cycling events, which cause cardiovascular volume overload. The event formulation is shown below, and some of the retrieved events are shown in figure \ref{fig:climb-event}
\newline
\begin{displaymath}
  UphillCycle := Cycling \wedge detect-climb(Altitude) \
\end{displaymath}

\begin{figure}[h]
  \centering
  \includegraphics[width=\linewidth]{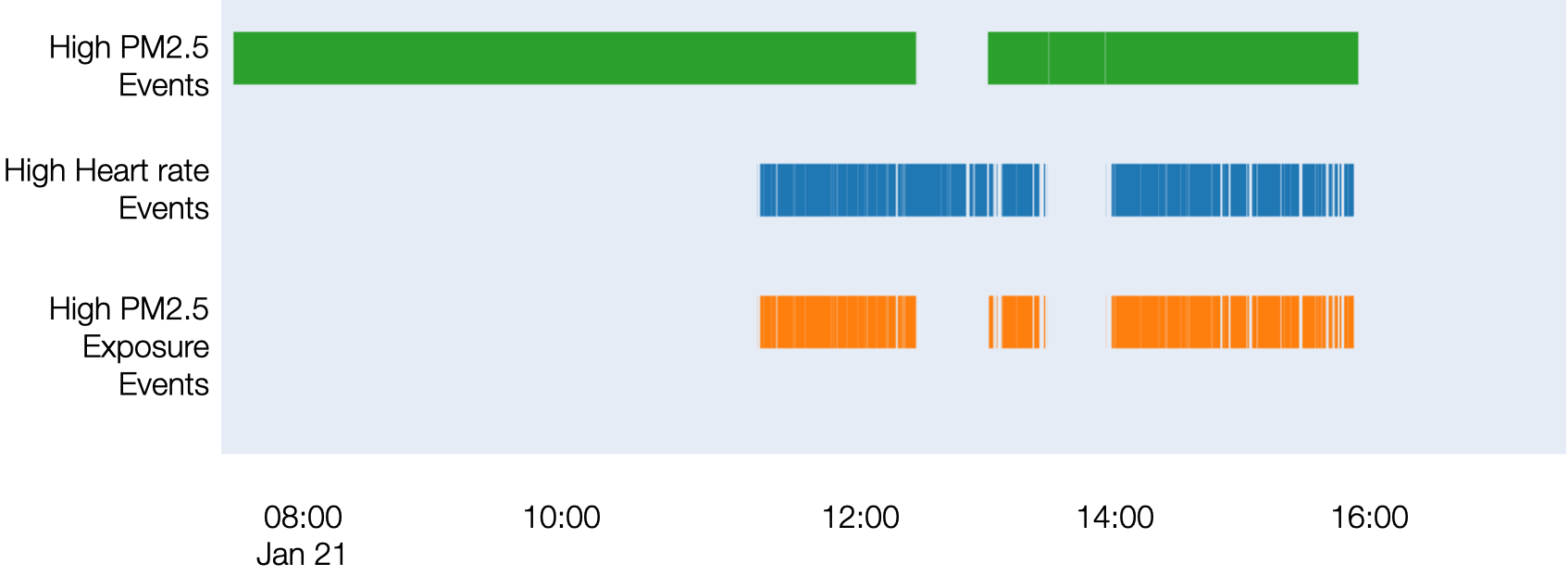}
  \caption{This figure depicts the instances during a day where the individual's PM2.5 intake is expected to be high. This is determined by combining heart rate zone events (HR > 120) with High PM2.5 concentration events, which in turn is determined using location stream and air pollution data.}
  \Description{PM2.5 Event Plot.}
  \label{fig:pm25-event}
\end{figure}

\begin{figure*}[h]
  \centering
  \includegraphics[width=\linewidth]{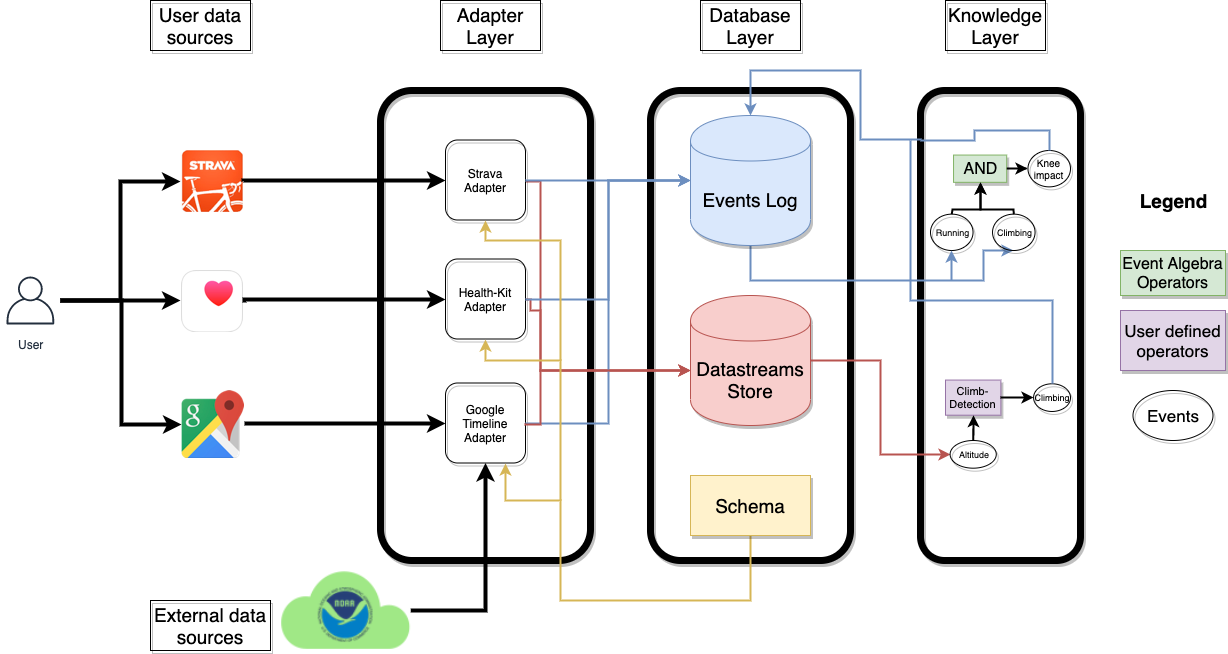}
  \caption{This figure shows a high level description of the system, starting from different data sources controlled by the user. These sources pass their data to the respective data adapters which convert the data to a unified schema. This schema is followed in the events log and the data stream store. The knowledge layer consists of definition of various complex events which are derived from measured events and data streams.}
  \Description{System Architecture.}
  \label{architecture}
\end{figure*}

\begin{table}
  \caption{Available sample schema for Data and event streams. Source adapters can choose a schema for different data streams being generated from that source. All events need to follow the event schema specified below. Different event attributes can be stored under the parameters field.}
  \label{tab:sample-schema}
  \begin{tabular}{cll}
    \toprule
    Stream & Field name & Data type\\
    \midrule
    Event & Event type & String\\
          & Event name & String\\
          & Start time & Timestamp\\
          & End time & Timestamp\\
          & Parameters & Key-Value pairs\\
          & Data streams & Set\\
          \\
    Real valued Data Stream & Timestamp & Timestamp\\
          & Value & Numeric\\
          & Unit & String\\
          & Source & String\\
          \\
    Location Data Stream & Timestamp & Timestamp\\
          & Value & Location\\
          & Unit & String\\
          & Source & String\\
  \bottomrule
\end{tabular}
\end{table}

\section{System Architecture}
As discussed in previous sections, there are two primary components of this system, and we will discuss their architecture separately. Event aggregation and knowledge-driven retrieval are very loosely coupled in this work so that the event retrieval framework could be used on different platforms, which need not follow the same data model.
\newline
Figure \ref{architecture} depicts a high-level architecture of the whole system and illustrates the flow of data between different components. 
\subsection{Event Aggregation}
Different data sources have their schema and modalities; thus, it is essential to have a separate parser for each source, which converts the incoming data to a uniform format and adds a reference to corresponding data streams for each event. The output of the adapters is different events measured by these systems and the associated data streams. All the adapters' output must follow one of the schemas depicted in table \ref{tab:sample-schema}.
The events log table contains, and each data stream has its table in the relational database. 
\newline
All data adapters have the same basic framework. They access a source-relation mapping to identify corresponding relations for every data stream and, if needed, create a new table for the data stream by picking one of the available schema (Fig. \ref{architecture}). After creating different relational data objects for the data streams and event streams, the adapters pass these objects to a database connector object which handles the I/O operations in the database. 
\newline
This data model allows us to create a new data stream by combining data from multiple sources, thus providing a less sparse data source for events. Data from different sources can also be combined to provide a more accurate estimate of the underlying value. We can also enrich existing events by using aggregated data streams.

\subsection{Event Operators and Transformations}
All event operators and transformation are implemented as functions that process a data stream or event stream, where event operators requiring more than one stream accept those as additional inputs. The different transformations are in different modules, and the appropriate module is selected based on the interface event definition provided.
The transformation functions can create a new data stream or a new event stream that can then be passed to the database connector for storing in the events log or data stream storage. Thus once a new stream has been created, it can be used like any other event or data stream for creating further events.
\newline
We expect this system to be used for estimating different biological variables in different contexts. The estimation framework should be able to query the system for relevant interface events. For example, if we are trying to estimate the cardiovascular health of an individual, the researcher or the user should be able to specify the interface events using the event pattern language. The estimation framework can trigger the interface events retrieval by querying for those, or it can be made continuous by adding appropriate triggers whenever a new event is added to the events log.

\section{Methodology}

In this paper, we present a system for retrieving interface events from multimodal data and event streams. We obtain the definition for different such events from medical literature and identify lifestyle events that could be utilized to retrieve those. For example, exposure to different pollutants is an important interface event which impacts different aspects of a person's health. We can calculate the pollutant exposure from the user's continuous GPS location stream and combining it with publicly available pollutant information provided by EPA \cite{DownloadEPA}.
\newline
Once the interface events have been defined, we can use the definition to retrieve those events and then utilize them for further estimation or guidance.

\subsection{Dataset}
We have used lifestyle and physiological data streams collected by an individual over ten years for our experiments. Different data streams have been collected over different periods and vary in the sampling frequency. We expect our system to be used for experiments in an N-of-1 setting \cite{Lillie2011TheMedicine} for providing individualized health guidance using observational data; thus, the evaluation of this system should be done for an individual over a period of time. We plan to instantiate the system for multiple users in future projects.
\newline
There are two primary sources of personal data that we have utilized in our experiments:
\begin{itemize}
    \item Detailed physiological data generated during an exercise event sampled at per second frequency. These data streams and events are collected from Strava and can provide a greater understanding of the user's health state.
    \item Lifestyle data collected as events or data streams during the course of the day. These are typically sampled at a lower frequency. These data streams and events are collected from various sources such as Apple HealthKit and Google timeline etc. These capture the lifestyle events that we expect to encounter during our day-to-day lives such as sleep, meal times, and commute.
\end{itemize}
These data streams and corresponding sources are listed in table \ref{tab:datasources}.
\newline
We have also used the pollution data captured by various environmental agencies across the world and aggregated by EPA. The data files can be found at \cite{NoTitleb}. We find the monitoring station nearest to the user's location and use the file generated by the station at the given time to estimate the user's exposure to different pollutants.
\begin{table}
  \caption{Data streams and sources}
  \label{tab:datasources}
  \begin{tabular}{cl}
    \toprule
    Data stream&Sources\\
    \midrule
    Heart rate & Strava, Apple Health-Kit\\
    Power & Strava\\
    Cadence & Strava\\
    Altitude & Strava\\
    Location & Strava, Google Location History\\
    Step Count & Apple Health-Kit\\
    Weight & Apple Health-Kit\\
    Stairs & Apple Health-Kit\\
    
  \bottomrule
\end{tabular}
\end{table}

\subsection{Interface Events}
We have defined interface events derived from biomedical literature using the previously mentioned event operators and specific event-detection operators for data streams.
We are retrieving two lifestyle based interface events and two environmental interface events which are described below.
\begin{itemize}
    \item \textbf{Lifestyle interface events} \textit{Volume Overload} events require an individual's heart to put a sustained effort. These can be identified from the heart rate stream (by finding intervals of high heart rate) or by combining altitude stream (to detect climb events) and cycling (or running, or hiking) events. If these events are repeated over time, they are likely to result in an increase in left ventricle volume (as depicted in fig. \ref{fig:heart}). These events can be represented as 
    \begin{displaymath}
      VolOverload := (HR > 140) \vee \\
      (Cycling \wedge detect-climb(Altitude))
    \end{displaymath}
    \textit{Pressure Overload} events require the heart to put in a short and intense effort, which presents as a spike in the heart rate stream. These can also be determined by using the power output stream collected during cycling events and identifying intervals where power output is higher than 400W. Over time, these events lead to a reduction in left ventricular volume. These events can be represented as
    \begin{displaymath}
      PressOverload := detect-spike(HR) \vee (Power > 400 W)
    \end{displaymath}

    \item \textbf{Environmental interface events} An individual's exposome streams can be created from their location history combined with publicly available GIS data provided by different government organizations. \textit{PM2.5 intake} can be computed from air pollutant data (provided by EPA) and the heart rate stream (used to estimate the breathing rate using results from \cite{Wasserman1973AnaerobicExercise}). We are estimating individual tidal volume using formulation given in \cite{Diacon2006Challenges10}. Multiplying the estimated tidal volume and breathing rate gives us the total air intake per minute for the individual, using which we can find their PM2.5 intake. Prolonged PM2.5 exposure has many long-term and short-term health effects\cite{Xing2016TheSystem}. \newline
    \textit{Blood $O_2$ level} of individuals is known to decrease with altitude as air pressure decreases with an increase in altitude. It leads to reduced blood oxygen saturation levels, which can be quantified by using the relationship described in \cite{HackettPHandRoachRCandSutton1995High-altitudeMedicine}. Low blood oxygen levels can lead to hypoxia, which can affect various biological functions \cite{Haase2006Hypoxia-inducibleKidney}. 
    
\end{itemize}



 
    

\begin{figure*}[h]
  \centering
  \includegraphics[width=\linewidth]{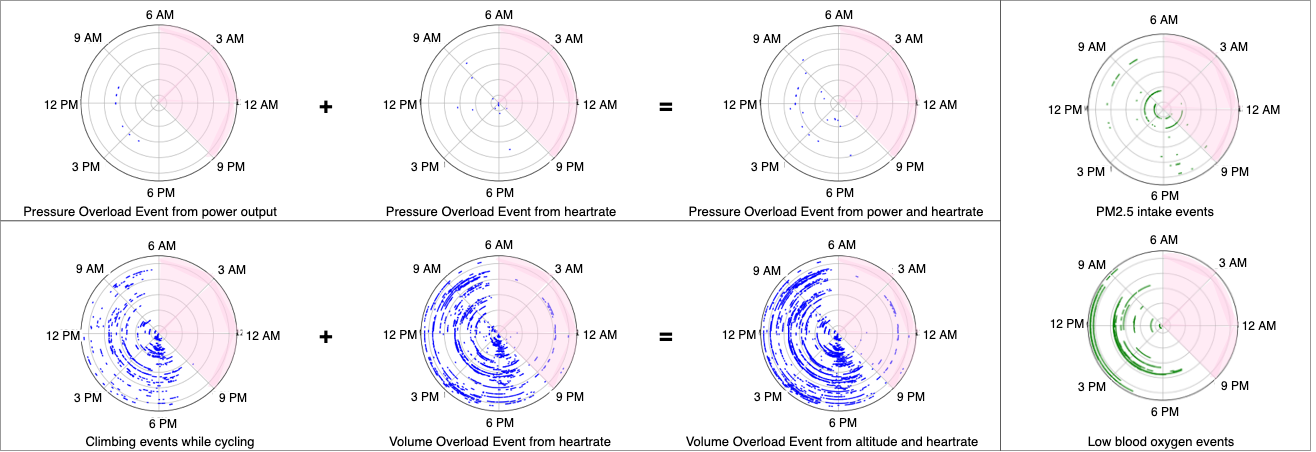}
  \caption{This figure shows different interface events retrieved using user-generated data. Different days of the year are represented as concentric circles, and different sectors of the circle represent different times of the day. Events in a day are represented as colored arcs on that circle. Volume and pressure overload events are retrieved using lifestyle data and events such as heart rate and exercise events. We can see that we can retrieve a significantly larger number of interface events by combining events from multiple data streams. This figure also shows the environmental interface events. High PM2.5 intake events are recognized over one week. These are the instances where per minute PM2.5 intake was higher than 0.7 $\mu$g. Low blood oxygen events are recognized over one year, where blood oxygen saturation goes below 95\%. The highlighted sectors roughly represent the time between sunrise and sunset; thus, we can see how different events overlap with circadian patterns.}
  \Description{Overload events.}
  \label{fig:overload-event}
\end{figure*}


\begin{figure}[h]
  \centering
  \includegraphics[width=\linewidth]{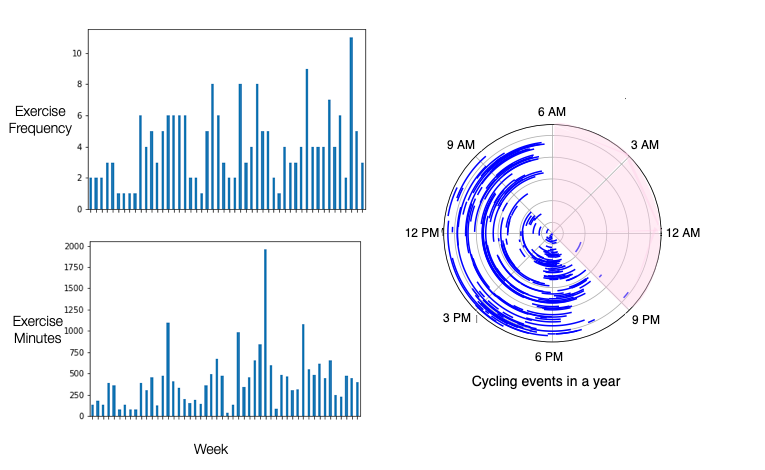}
  \caption{This figure shows the results of a real-world query about exercise behavior. The bar plots show the exercise frequency and exercise minutes per week over a year. The polar plot shows all the exercise events during the day, where the highlighted sector roughly matches the time between sunset and sunrise. Thus we can see how likely are these events to cause any disruptions in circadian patterns.}
  \Description{Real world events.}
  \label{fig:real-query}
\end{figure}

\section{Results}

\subsection{RQ1: Data fusion}
We can use multiple data streams from different sources to recognize different instances of an interface event. This is evident in fig. \ref{fig:overload-event}, which depicts the occurrences of cardiovascular volume and pressure overload events. Volume overload events are detected in two ways, 1) We identify intervals with a sufficiently high heart rate, and 2) We identify intervals where the person is climbing up a slope under their power (e.g., cycling). These two event definitions are combined using an $OR$ operator to give all occurrences of volume overload events. We could add more methods of recognizing volume overload events using the same operator with minimal additional effort. 
Similarly, pressure overload events are being recognized from heart rate data stream (spike detection) and power stream (high effort intervals). These events are, by definition, short, and the same can also be seen in fig \ref{fig:overload-event}. We can also combine the user-generated data with environmental data to generate their exposome streams. Figure \ref{fig:overload-event} shows the occurrences of PM2.5 intake events over a week, and low blood oxygen events over a year. 

\subsection{RQ2: Continuous Retrieval}
\begin{figure}[h]
  \centering
  \includegraphics[width=\linewidth]{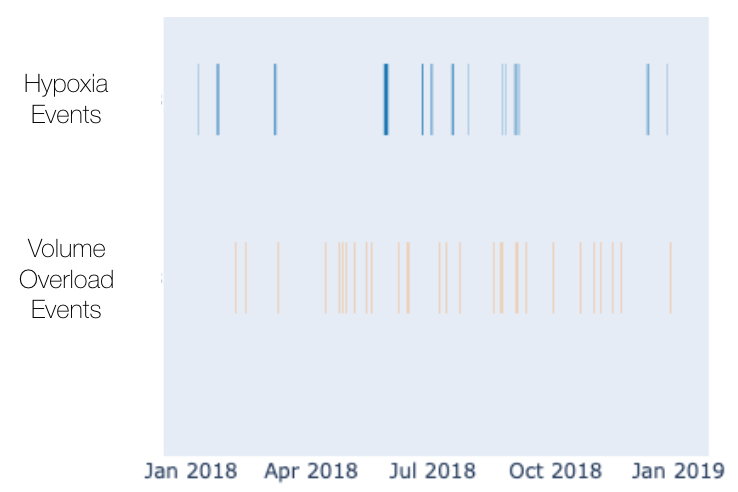}
  \caption{This figure shows the continuous retrieval of volume overload and hypoxia events over the course of a year.}
  \Description{Continuous retrieval.}
  \label{fig:continuous-retrieval}
\end{figure}

We are retrieving the events continuously over a year, as shown in figure \ref{fig:continuous-retrieval}. A zoomed-in view is shown in figures \ref{fig:climb-event} and \ref{fig:pm25-event} which highlight the combinations of events while retrieving an interface event. We apply event detection operators on incoming data streams and use the definition of the interface events to combine the generated events. Once we have retrieved the interface events over a sufficiently long period, we can analyze their occurrences and try to find useful patterns, such as relationship of volume overload events with circadian patterns. We can see in fig. \ref{fig:overload-event}, the density of the volume overload events is very low after sunset, which may lead us to think that any disruptions in sleep patterns are unlikely to be caused by these events. We can overlay a similar plot of sleep events on this plot and see if there is any relationship between the event streams.

\subsection{RQ3: Real world query}
The system is also useful for answering a real-world query about a person's life in terms of attributes and patterns of relevant interface events. One such example is in fig. \ref{fig:real-query} where we try to respond to a physician's questions about a person's exercise habits. Typically a doctor would ask the person about the frequency and intensity of their exercise routine. These answers are in figure \ref{fig:real-query}, where we display the exercise frequency and total duration of exercise events for every week of the year. We also show all the exercise (cycling) events in a format that can be compared with any other lifestyle habit (e.g., sleep) to facilitate the discovery of new behavioral/physiological patterns.


\begin{figure}[h]
  \centering
  \includegraphics[width=\linewidth]{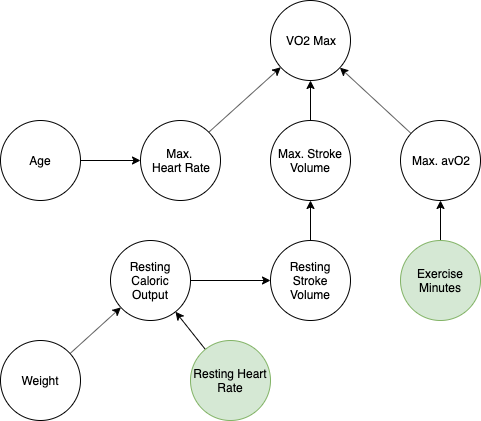}
  \caption{This figure shows the relationship between different variables required for estimating VO2 Max from daily life data and event streams. The highlighted nodes refer to interface events or data streams created from interface events.}
  \Description{Estimation Graph.}
  \label{fig:graph}
\end{figure}

\section{Conclusion}
In this paper, we propose an approach for specifying and retrieving health events, which we call interface events, using the domain knowledge from biomedical literature. As we showed, using multiple data streams in a knowledge-driven manner allows us to leverage the available medical literature and also leaves the possibility of adding new data streams and information sources in the future. One drawback of this approach is the effort needed to convert the existing knowledge into event operator language. However, we believe that with the advances in natural language processing and knowledge graphs, we may be able to automate this process by directly converting the insights from medical literature to events definition. 
The approach detailed in this paper facilitates continuous health state estimation and, eventually, continuous health navigation, as described in \cite{Nag2019ALife}.
\subsection{Application: Health State Estimation}
We have used the event retrieval framework in conjunction with a health state estimation framework, the details of which are beyond the scope of this work. We are trying to estimate the cardio-respiratory fitness of the user using $VO_2 Max$ as an indicator. The graph depicting the causal relationship between various measured and estimated variables is depicted in fig. \ref{fig:graph}.
The interface events for this estimation scheme were sleep and exercise events. 
We used sleep events to estimate the user's resting heart rate and aggregated the exercise events' duration to estimate the change in arteriovenous oxygen difference from its base value \cite{ChapterCDC}. The details for this project could be found in \cite{Nag2020HealthEstimation}.
\newline
A nutrition guidance/recommendation system could utilize such a system to calculate an individual's electrolyte requirements by combining their location stream with local weather to estimate their activity levels and local temperature, which in turn allows us to estimate how much sodium they lost through sweat\cite{Nag2017LiveEngine}. This can be generalized to other macronutrients too. 
\subsection{Future Work: Knowledge Discovery}
We can employ knowledge discovery algorithms to find recurring patterns between interface events and changes in biological variables or symptoms which do not have a well-defined relationship with each other. For example, we could discover if a person is allergic to a particular pollutant (e.g., pollen) by correlating their exposome stream with adverse health outcomes.
Causal discovery algorithms could be used to find a causal link between different lifestyle events and biological parameters\cite{CaiCausalRepresentation}. Such causal links indicate an unknown interface event. Situations like this would need to be explored further by the experts in that particular domain.
\newline
We plan to build upon the existing system and discover relationships between lifestyle events and biological parameters.
\bibliographystyle{ACM-Reference-Format}
\bibliography{references}

\end{document}